\documentclass[12pt,preprint]{aastex}
\begin{document} 


\title{The True Incidence of Magnetism Among Field White Dwarfs} 

\author{James Liebert}
\affil{Steward Observatory, University of Arizona, Tucson, 
AZ 85721.}
\email{liebert@as.arizona.edu} 
\author{P. Bergeron}
\affil{D\'epartement de Physique, Universit\'e de Montr\'eal, C.P.~6128, 
Succ.~Centre-Ville, 
Montr\'eal, Qu\'ebec, Canada, H3C 3J7.}
\email{bergeron@astro.umontreal.ca}
\and
\author{J. B. Holberg}
\affil{Lunar and Planetary Lab, University of Arizona, 
Tucson, AZ 85721.}
\email{holberg@argus.lpl.arizona.edu}

\begin{abstract} 

We study the incidence of magnetism in white dwarfs from three large and
well-observed samples of hot, cool and nearby white dwarfs in order to
test whether the fraction of magnetic degenerates is biased, and whether
it varies with effective temperature, cooling age, or distance.  The
magnetic fraction is considerably higher for the cool sample of
Bergeron, Ruiz, \& Leggett, and the Holberg, Oswalt, \& Sion sample of
local white dwarfs than it is for the generally-hotter white dwarfs of
the Palomar Green Survey.  We show that the mean mass of magnetic white
dwarfs in this survey is 0.93~$M_{\odot}$ or more, so there may be a
strong bias {\it against} their selection in the magnitude-limited
Palomar Green Survey.  We argue that this bias is not as important in
the samples of cool and nearby white dwarfs.  However, this bias may not
account for all of the difference in the magnetic fractions of these
samples.

It is not clear that the magnetic white dwarfs in the cool and local
samples are drawn from the same population as the hotter PG stars.  In
particular, two or three of the cool sample are low-mass white dwarfs in
unresolved binary systems.  Moreover, there is a suggestion from the
local sample that the fractional incidence may increase with decreasing
temperature, luminosity, and/or cooling age.  Overall, the true
incidence of magnetism at the level of $\sim$2~MG or greater is at least
$\sim$10\%, and it could be higher.  Limited studies capable of
detecting lower field strengths down to $\sim$10~kG suggest by
implication that the total fraction may be substantially higher then
10\%.

\end{abstract} 


\section{Introduction}

For three decades it has been known that a small minority of white
dwarfs (WDs) have detectable magnetic fields, but the percentage has
remained somewhat uncertain.  Angel, Borra, \& Landstreet (1981)
summarized the early surveys with the statement: ``We find that BP(B),
the probability per octave, is roughly constant at 0.005 for fields in
the range 3$\times$10$^6$~G to 3$\times$10$^8$~G, and does not exceed
this value down to 10$^6$~G.''  Below 3$\times$10$^6$~G they noted that
more precise longitudinal polarization measurements were necessary to
detect magnetism.

With the introduction of CCD detectors more than a decade ago, surface
field strengths well below 10$^6$~G were accessed for the first time.
Schmidt \& Smith (1994, 1995) reported results from extensive surveys
using a polarimeter and spectropolarimeter.  In the first reference,
they concluded ``The overall incidence of magnetism among white dwarfs
remains low, with more than 90\% of stars having fields below $\sim$10
kG[kilogauss].''  In the second reference they stated ``The incidence of
magnetism among white dwarfs is found to be $4.0\pm$1.5\% for fields
between 3$\times$10$^4$ and 10$^9$~G.''

In their comprehensive review paper describing the properties of the 65
magnetic white dwarfs (MWDs) then known, Wickramasinghe \& Ferrario
(2000, hereafter WF00) had this summary statement: ``The isolated MWDs
comprise $\sim$5\% of all WDs.''  Such modest fractional estimates have
led several authors, perhaps beginning with Angel et al. (1981) and
later WF00, to hypothesize that the MWDs are the progeny of the peculiar
A-B stars, the only main sequence stars known to have formed with
``fossil'' magnetic fields emanating from the stellar cores.

Evidence that the incidence of magnetism could be higher at least for
particular kinds of WDs has also been noted.  For example, WF00 noted
that about 25\% of known cataclysmic variables harbor MWDs as
primaries, but they discuss selection effects in favor of discovery
of, for example, the more X-ray luminous AM~Herculis and DQ~Herculis
accreting MWD systems.  Hence, it remains uncertain whether the
fraction of MWDs is really higher for a true, volume-limited sample of
CVs vs. the isolated WDs.

Vennes (1999) found that the fraction of MWDs is higher (about 25\%)
among stars with masses exceeding 1~$M_{\odot}$ in his Extreme
Ultraviolet Explorer (EUVE) sample of hot white dwarfs.  This is
consistent with studies of individual cooler stars (e.g., Greenstein \&
Oke 1982; Liebert 1988) that many of the known MWDs have quite high
masses (see also the discussion by WF00).

Liebert \& Sion (1978) noted decades ago that the magnetic field
discovery fraction for isolated WDs was higher for the cooler than for
the hotter WDs.  It should also be noted that the cooler stars are
generally nearer to the Sun than are the hotter WDs found in UV-excess
or blue color surveys.  Fabrika \& Valyavin (1999) studied published
results and conclude that the frequency of magnetism among hot WDs is
$3.5\pm0.5$\%, but $\gtrsim 20\pm5$\% for white dwarfs cooler than
10,000~K.  They note that most of the known cool MWDs are within a
distance of 25~pc. Holberg, Oswalt, \& Sion (2002) tabulated and studied
the sample of known white dwarfs within 20~pc distance.  Kawka et
al. (2002) already have noted that the percentage of MWDs in this sample
exceeds 10\%; this volume-limited sample is of course dominated by cool
WDs.
 
Theoretical expectation, however, is that fossil magnetic fields in WDs
should not increase in strength, but rather should decay slowly
(cf. Fontaine, Thomas, \& Van Horn 1973; see, however, Valyavin \&
Fabrika 1999 for a different perspective). It must be admitted, in any
case, that different discovery techniques and biases are involved in the
discovery of hot and cool WDs.  WF00 concluded that there is no
convincing evidence that the mean magnetic field strength varies with
$T_{\rm eff}$, but they did not comment on the incidence of detected
magnetism.

One problem has been that existing data sets are quite heterogeneous,
represent different WD discovery techniques, and varying precision in
the search for magnetism.  In this paper we revisit these questions,
beginning first in Section 2 with new statistics for three large and
well-studied samples of cool and hot white dwarfs now available.  In
Section 3, we attempt to sort out the sources of the apparently-
conflicting results, to reach a more robust conclusion about the
true fraction of MWDs, and whether it changes as the white dwarfs 
cool.  

\section{Comparison of Hot, Cool and Nearby White Dwarf Samples}

\subsection{Hot White Dwarfs from the Palomar Green (PG) Survey} 

Liebert et al. (2003) have been finishing a detailed study of 324 DA
white dwarfs from the PG with $T_{\rm eff}$ above about 10,000~K
including high signal-to-noise ratio optical CCD spectra covering all
Balmer lines except H$\alpha$.  The technique of simultaneous
line-fitting for the determination of $T_{\rm eff}$, surface gravity
($\log g$), combined with the use of evolutionary tracks to determine
the mass, radius and the cooling time, is now well-established
(cf. Bergeron, Saffer, \& Liebert 1992).  Except for the very hottest
objects (perhaps T$_{eff} \gtrsim$ 50,000~K or even higher), the precise
measurement of line profiles makes possible the discovery of any
magnetic fields down to a limit of $\sim$2 megagauss (MG), below which
clean Zeeman-splitting of H~I and He~I does not occur.  It is possible
that careful measurements of the positions of the high Balmer lines
could detect quadratic Zeeman shifts to significantly lower field
strengths (Preston 1970), but our more conservative assessment here is
to assume that this has not been done systematically with the spectra.
Seven DA MWDs have been found from this sample, and these are listed in
order of decreasing surface field strength in Table~1.

Some comments on individual objects are in order.  For the parameters of
PG~2329+267, we favor the fitted values from Bergeron, Leggett \& Ruiz
(2001, hereafter BLR -- see next section) over the rough mass estimate
of 0.9~$M_\odot$ based on inspection of spectra by Moran, Marsh, \&
Dhillon (1998).  Revised parameter estimates from our study are listed
for PG~1220+234, and the fit to the spectrum is shown as Figure~1.  A
temperature of 14,000~K has been assumed for PG~1015+015, based on the
spectrum fit of Wickramasinghe \& Cropper (1998).  This star has too
strong a field for the line profile-fitting to yield an estimate of the
surface gravity.  However, an unpublished trigonometric parallax for
this star allows a rough mass estimate of 1.15~$M_\odot$.

Similar studies cover 24 PG~DB white dwarfs by Beauchamp (1996) and
include eight PG~DO stars by Dreizler \& Werner (1996).  One magnetic
PG~DB star was found from spectropolarimetric observations (Putney 1997,
Wesemael et al. 2001); it is also listed in the table.  At an estimated
polar field strength of 1~MG, however, its field was not detected in the
spectrophotomeric study.  Finally, the table includes two magnetic stars
cooler than 10,000~K -- G195$-$19 and G128$-$72 -- whose fields were
detected in other studies, but also included in the PG survey.  The
catalog included 22 other DC white dwarfs, two DZ, and eight cool DA
white dwarfs in the survey for which modern spectra may not be
available.  Since the stars cooler than 10,000~K are not part of the
comprehensive spectrophotometric data set, they are not included in the
assessment of the magnetic frequency of the hot PG stars.  However, both
G195$-$19 and G128$-$72 are included in the cool star sample discussed
next.

\begin{deluxetable}{llrrrr}

\tablenum{1}
\tablecaption{ Magnetic White Dwarfs in the Palomar Green Survey }
\tablewidth{0pt}
\tablehead{ 
\colhead{PG} & \colhead{Other Name} & \colhead{B$_p$(MG)} &
\colhead{Comp} & \colhead{Mass ($M_\odot$)} & \colhead{$T_{\rm eff}$ (K)} }

\startdata

1031+234 & & 1000: & H & 0.93 & 21,700 \\
0945+245 & LB~11146 & 670 & H,He & 0.93 & 14,500 \\
0912+536 & G195$-$19 & 100: & He & 0.75 & 7160 \\
1015+015 &  & 90 & H & 1.15 & 14,000 \\
1533$-$057 &  & 31 & H & 0.94 & 20,000 \\ 
1312+098 &  & 10 & H & & 15,000 \\ 
1658+441 &  & 3.5 & H & 1.31 & 30,500 \\
1220+234 &  & 3: & H & 0.81 & 26,500 \\
2329+267 & G128$-$72 & 2.3 & H & 0.61 & 9400 \\ 
0853+164 & LB~8827 & 1: & He,H & & 23,000 \\

\enddata
\end{deluxetable}

The bottom line is that only seven MWDs with field strengths $\ga$2~MG
were found from 356 well-observed stars, or $2.0\pm0.8$\%.  Despite the
homogeneous spectrophotometric studies, it should be noted that no
special polarimetric or spectropolarimetric study has focussed on the PG
sample, though PG~0853+164 (LB~8827) happened to be targetted.

Another interesting property of the PG subset of MWDs is the
distribution in mass, a parameter available for all but one with
$\ga$2~MG field strengths (but we do include the two cooler magnetic 
stars).  The mean mass is 0.93~$M_\odot$, compared
with a value near 0.58~$M_\odot$ for the hot DA (Liebert et al.  2003,
in preparation) and DB samples (Beauchamp 1996).  We shall discuss the
implications of these values in Section~3.

\subsection{Cool White Dwarfs} 

Extensive studies of known cool white dwarfs have been published by
Bergeron, Ruiz, \& Leggett (1997, hereafter BRL) and BLR.  BRL included
111 white dwarfs cooler than about 10,000~K, but selection emphasized
non-DA spectral types at the expense of DAs.  BLR studied 152 cool white
dwarfs of all spectral types which had published, accurate trigonometric
parallaxes from which luminosities, masses and radii (in addition to
$T_{\rm eff}$) could be estimated.  All but a few have $T_{\rm eff}$
estimates below 10,000~K.  There is considerable overlap in these two
samples, so they are combined in our analysis, a total of 199 individual
WDs (hereafter, the BRL sample).

\begin{deluxetable}{llrlrr}

\tablenum{2}
\tablecaption{ Magnetic White Dwarfs from BRL and BLR}
\tablewidth{0pt}
\tablehead{ 
\colhead{WD} & \colhead{Other Name} & \colhead{B$_p$(MG)} &
\colhead{Comp} & Mass ($M_\odot$) & \colhead{$T_{\rm eff}$ (K)} }

\startdata

1900+705 & Grw+70~8247 & 320 & H & 0.95 & 12,070 \\
1829+547 & G227$-$35 & 175 & H & 0.90 & 6280  \\
1036$-$204 & LP~790$-$29 & 100: & C$_2$(He) & & 7500 \\
0912+536 & G195$-$19& 100: & He & 0.75 & 7160 \\
1748+708 & G240$-$72 & 200: & He & 0.81 & 5590 \\ 
0553+053 & G99$-$47 & 25 & H & 0.71 & 5790 \\ 
1026+117 & LHS~2273 & 18 & H & & 7160 \\
0011$-$134 & LHS~1044 & 16.7 & H & 0.71 & 6010  \\
1639+537 & GD~356 & 13 & H & 0.67 & 7510 \\ 
0548$-$001 & G99$-$37 & 10-20 & C$_2$(He) & 0.78 & 6400 \\
1330+015 & G62$-$46b & 7.4 & H & 0.25 & 6040 \\
0503$-$174 & LHS~1734 & 7.3 & H & 0.32 & 5300 \\
2329+267 & G128$-$72 & 2.3 & H & 0.61 & 9400 \\
1503$-$070 & GD~175 & 2.3 & H & 0.70 & 6990 \\
1953$-$011 & LHS~3501 & $\lesssim$0.6: & H & 0.74 & 7920 \\
0009+501 & G217$-$37 & $\lesssim$0.2 & H & 0.74 & 6540 \\
0257+080 & LHS~5064 & $\lesssim$0.2 & H & 0.57 & 6680 \\

\enddata
\end{deluxetable}

The MWDs found in the BRL sample are listed in Table 2.  Before further
discussion, a few corrections to the tabulation of WF00 need to be
noted.  The spectrophotometric discovery of magnetism in GD~175
(1503$-$070) was reported in BLR, and was not available for WF00.
However, there are four stars from the BRL sample listed in WF00 which
we do not believe to be magnetic.  G141$-$2 (1818+126) and G227$-$28
(1820+609) were shown by Putney (1997) {\it not} to have detectable
circular polarization.  The same conclusion for ESO~439$-$162
(1127$-$311A) was reached by Schmidt, Bergeron, \& Fegley (1995); WF00
considered it to be a possible MWD.

Finally, there is the case of LHS~1008 (0000$-$345).  Based on what was
reported to be a 5-minute exposure at the ESO 1.5-m telescope, Reimers
et al. (1996) reported a ``complex H$\beta$'' feature near 4650 \AA, and
fitted a model with H$\alpha$ Zeeman splittings.  This resulted in an
estimated field strength of 86~MG.  This value is listed in WF00.  Our
spectrophotometry, obtained with the CTIO 4-m Blanco reflector, was
analyzed in BRL.  This showed a DC spectrum, with no features at 4650
\AA\ or H$\alpha$.  Even stronger evidence that the star exhibits no
magnetic field was provided by the spectropolarimetric analysis of
Schmidt et al. (2001), which also shows that the spectrum is cleanly DC.
We thus exclude this object from further consideration as a MWD.
 
The BRL sample yields a fractional incidence of 14/199 or $7.0\pm2$\%
of MWDs with field strengths $\ga$2~MG, formally more than three times
the fraction found for the PG sample.  This does not include the three
stars listed in Table~2 with lower or uncertain field strengths.
 
Mass estimates are available for all but two $\ga$2~MG MWDs (Table 2).
The mean mass is 0.68~$M_\odot$.  However, these mass estimates are based
on a different method than the direct determinations of the surface
gravity.  BRL and BLR estimated masses by fitting the absolute magnitudes
derived from available trigonometric parallaxes.  The evidence is
mounting, however, that a fraction of WDs exceeding 10\% may be
unresolved binaries (Napiwotzki et al. 2002), and the apparent magnitude
difference of the components may be expected to grow smaller as the
stars cool.  If the fitted luminosity is due to two, rather than one
star, the inferred mass of the ``single'' star may be underestimated
drastically, as noted in both BRL and BLR.  It is therefore possible 
that double degenerates in these samples have skewed the mean mass
estimate downwards.  Moreover, the lack of precision of this method of
estimating masses may smear out the expected peak of the mass
distribution near 0.6 $M_\odot$, the mean found for samples analyzed by
the gravity-fitting method (e.g., Bergeron, Saffer, \& Liebert 1992). 

In fact there are three definite or possible binaries for which one
component is a {\it low} mass MWD of DA type.  G62$-$46 was shown to be
composite spectrum object consisting of the DA ``b'' component whose
parameters are listed in Table~2, with a warmer, more massive DC ``a''
component (Bergeron, Ruiz, \& Leggett 1993).  The main reason for
invoking a binary hypothesis is that the H$\alpha$ Zeeman absorption
components are too weak for a DA of the $T_{\rm eff}$ derived by fitting
the energy distribution under the assumption that the flux is from a
single star.  These authors discussed the possibility that LHS~2273,
which also exhibits H$\alpha$ components too weak for the energy
distribution, could be decomposed into a binary with similar properties
(however, the parameters listed in Table~2 are taken from the single WD
fit in BRL).  The third such candidate is LHS~1734, an apparently-single
magnetic DA of similar $T_{\rm eff}$ to the other two DA components, but
for which the parallax (luminosity) requires the assignment of a very
low mass of 0.32~$M_\odot$.  Such low mass WDs should never have reached
the required mass for core helium ignition, and should be composed of
helium interiors.  It is difficult to understand the evolution to the
helium white dwarf stage within the lifetime of the Galaxy except to
invoke binary mass exchange.  LHS~1734 differs from the other two in
that the H$\alpha$ line strengths are in agreement with the 5300~K fit
to the energy distribution.  That is, any companion that we argue must
have engaged earlier in close-binary mass exchange must not be
contributing significant light at red wavelengths.
 
That three of these 14 MWDS have such low masses may certainly skew the
mean mass of the BRL sample.  Apparently, magnetic white dwarfs may form
in binary evolution in a different manner than for single stars, and in
particular may have very different mean masses.  It is also curious that
all have temperatures around 5-6000~K, as was discussed earlier in BRL.
Such binaries involving a low-mass MWD for whatever reason appear not to
be represented in the PG sample (although low-mass non-magnetic DAs are
certainly present there).  Without the two known binaries, the mean mass
of the MWDS in the BRL sample is 0.76~$M_\odot$.

\subsection{The Local Sample} 

As noted earlier, Holberg et al. (2002) tabulated 107 known WDs whose
best distance estimates indicate that they are within 20 parsecs.  Since
two of these have been determined to be binary white dwarfs for which we
list parameters of the individual components, our sample consists of 109
stars.  Unlike the two samples considered previously, this sample lacks
a homogeneous spectrophotometric data set, although a majority are in
the BLR sample.  These authors argue that the sample is complete for the
46 stars out to a distance of 13 parsecs, though the completeness falls
to 65\% at 20~pc.  For reasons discussed in the previous section, we
exclude LHS~1008 and G227$-$28 from the subset of local MWDs, which are
listed in Table~3.  Distance estimates are listed in the last column.

As Kawka et al. (2002) have independently noted, this sample also has a
high yield of MWDs.  Nine MWDs with field strengths $\ga$2~MG are listed
from the local sample, five with estimated distances within 13~pc.  The
fractions are 5/45 or $11\pm5$\% (13~pc) and 9/109 or $8\pm3$\% from the
local sample out to 20 pc.  Not included in the statistics are two stars
within 13~pc with weaker field strength estimates from Schmidt \& Smith
(1995).  The local MWD sample is almost a subset of the BRL sample --
all but G256$-$7 (1309+853) were analyzed in those studies.

Mass estimates are available for all but two MWDs from the local sample
with field strengths $\ga$ 2~MG.  The mean is 0.80~$M_\odot$.  Since the
mass estimates are again derived from trigonometric parallaxes and not
spectroscopic measurements of the surface gravity, this value is subject
to skewing by any unresolved binaries, as discussed in the previous
subsection.  However, there is no direct evidence that any of the MWDs 
are in binary systems. 

\begin{deluxetable}{llclcrr}

\tablenum{3}
\tablecaption{ Magnetic White Dwarfs within 20~pc }
\tablewidth{0pt}
\tablehead{ 
\colhead{WD} & \colhead{Other Name} & \colhead{B$_p$(MG)} &
\colhead{Comp} & \colhead{Mass ($M_\odot$)} & \colhead{$T_{\rm eff}$ (K)} &
\colhead{D(pc)} }

\startdata

1900+705 & Grw+70$^o$~8247 & 320 & H & 0.95 & 12,070 & 12.98 \\
1748+708 & G240$-$72 & 200: & He & 0.81 & 5590 & 6.07  \\
1829+547 & G227$-$35 & 175 & H & 0.90 & 6280 & 14.97 \\
1036$-$204 & LP~790$-$29 & 100: & C$_2$(He) & & 7500 & 13.27 \\ 
0912+536 & G195$-$19 & 100: & He & 0.75 & 7160 & 10.27 \\
0553+053 & G99$-$47 & 25 & H & 0.71 & 5790 & 7.95 \\ 
0011$-$134 & LHS~1044 & 16.7 & H & 0.71 & 6010 & 19.5 \\ 
0548$-$001 & G99$-$37 & 10-20 & C$_2$(He) & 0.78 & 6400 & 11.09 \\
1309+853 & G256$-$7 & 4.9 & & 0.50 & 5200: & 18.05 \\ 
1953$-$011 & LHS~3501 & $\lesssim$0.6 & H & 0.74 & 7920 & 11.39 \\ 
0009+501 & G217$-$37 & $\lesssim$0.2 & & 0.74 & 6540 & 11.04 \\

\enddata
\end{deluxetable}

\section{Discussion} 

\subsection{Correcting for the radius bias} 

How are these apparently-conflicting results concerning the incidence of
WD magnetism to be reconciled?  We believe that the answer lies first in
considering the high mean masses found for MWDs in all three samples
considered here (omitting the known binaries in BRL).  These results are
consistent at least qualitatively with the large fraction of massive,
EUV-selected white dwarfs reported by Marsh et al. (1997) and Vennes
(1999).

Since the PG Survey is explicitly magnitude-limited (approximately to
$B=16.2$), it could be subject to a large bias {\it against} the
inclusion of MWDs. As already noted by Liebert \& Bergeron (1994), the
survey limiting distance is proportional to the stellar radius ($R$),
and hence the survey volume for a given mass scales as $R^3$.  Thus the
fractional incidence of MWDs could be substantially smaller in the PG
Survey than in a true, volume-limited sample.  Other hot white dwarf
surveys (i.e. the Hamburg Schmidt) must be vulnerable to the same bias,
as is the magnitude-limited polarimetric survey of Schmidt \& Smith
(1995).  Ironically, similar EUV/X-ray magnitude limits apply to EUVE
and ROSAT detections, making the discoveries of many massive, magnetic
WDs by Vennes (1999) and Marsh et al. (1997) all the more remarkable.
The greater likelihood that massive WDs would have close to
pure-hydrogen atmospheres -- and correspondingly less EUV/soft X-ray
opacities due to helium and heavier elements -- may offset the
radius bias.

With the samples of ``local'' white dwarfs, the radius bias is more
difficult to quantify even if we knew the underlying mass distributions
of WDs and MWDs.  Since they are not explicitly magnitude-limited, the
bias is likely to be less important. Note that Holberg et al. (2002)
claim a complete sample to 13~pc.  The considerably-overlapping BLR
sample probably is also closer to volume-limited than is the PG.  

If one does have a complete sample to a given limiting magnitude, 
as we argue for the PG, one could correct for the difference in 
search volume for the MWDs if one knew their true distribution in 
mass.  However, we do not know this distribution.  Arguably, the 
PG MWD mass determinations are more accurate, since they are not 
subject to the binary bias.    

To estimate the potential magnitude of the mass-bias on the MWD
fractional incidence in the PG Survey, let us perform the following
exercise:  Assume that the true MWD distribution has the estimated mean
mass of 0.93~$M_\odot$, and that it otherwise has the same ``shape'' as
the non-magnetic mass distribution.  The typical PG star at the mass
distribution peak of 0.58~$M_\odot$ (and $\sim$25,000~K) has a radius of
0.0144~$R_\odot$, compared with 0.0091~$R_\odot$ for a mass of
0.98~$M_\odot$.  To estimate their representation within a fixed search
volume, each star should be assigned a weight proportional to $R^{-3}$.
This ratio leads to a properly-weighted relative fraction of MWDs of
2.0$\pm$0.8 $\times$ (0.0144/0.0091)$^3$ = 7.9$\pm$3\%.

This exercise brings the MWD fraction of the PG into better agreement
with the cool and nearby WD samples, although it necessarily has
substantial errors due to the small, intrinsic number of MWDs.  However,
there may still be uncorrected bias.  The mean mass of MWDs in the PG
could be biased downwards from a true mean mass, since a massive MWD has
less chance to be part of the survey than a lower mass MWD in the first
place.  Thus the relative fraction MWDs may be significantly higher
still, perhaps approaching 10\%.  Finally, there is also no rigorous way
of correcting this fraction upwards for the MWDs with field strengths
below $\sim$2~MG, which generally require spectropolarimetric
observations.

\subsection{Is there evidence for evolution in the incidence of 
large fields?}

The exercise of the previous paragraphs suggests that there may be no
significant difference between the fractional incidence of MWDs between
the hot and cool samples of WDs, once the greater correction for bias in
the magnitude-limited PG is allowed for.  This would enable us to retain
the simpler working hypothesis that the incidence of MWDs or mean
magnetic field strengths do not increase as the WDs cool.  It is not
surprising that the majority of the local sample are cool WDs.
Arguably, Liebert \& Sion (1978) were simply detecting the greater MWD
frequency of the more volume-complete sample of solar WD neighbors, as
do Holberg, Oswalt, \& Sion (2002).  However, one final test is
possible: if the local WD sample is close to being complete, is there
any significant difference between the $T_{\rm eff}$ distribution of the
MWDs vs the sample as a whole?

To perform this test, we have collected $T_{\rm eff}$ and mass estimates
for nearly all of the local sample, although we concede that this may
not be complete for stars more distant than 13~pc.  Excellent parameters
are available for the majority (63) which were analyzed in
BLR. Parameters for several more stars were found in Bergeron et
al. (1990), Bergeron et al. (1992), Bergeron, Liebert, \& Fulbright
(1995a), Bragaglia, Renzini, \& Bergeron (1995), Bergeron et
al. (1995c), Holberg et al. (1998), Koester et al. (2001), Provencal et
al. (2002), and also from a sample of DA spectra kindly made available
to us by C.~Moran (private communication) and analyzed by
us. Temperatures for two stars were also taken from BRL but with no mass
estimate.  This left 12 stars out of 108 without parameter estimates.
For these, we estimated the temperatures based on the published color
indices from the McCook \& Sion (1999) catalog and the photometric
calibration of Bergeron, Wesemael, \& Beauchamp (1995b); no mass
estimates are possible for these. Finally, two stars remain without even
sufficient information for McCook \& Sion (1999) to have estimated a
subtype.  These are omitted from this exercise.

In Figure~2 we plot the estimated $T_{\rm eff}$ values vs. mass for the
local sample.  Stars with no mass estimate, generally those whose
$T_{\rm eff}$ is taken from the spectral subtype, are plotted at the
bottom (at mass of 0.1).  Eleven of the 69 stars (16$\pm$5\%) cooler
than 8,000~K are magnetic, while only one of 32 WDs (3$\pm$3\%) hotter
than this temperature is a MWD.  This is therefore only about a two
sigma difference.  For the majority with mass estimates, we also can
evaluate the fraction vs. evolution time, and the isochrones are
displayed in the figure.  Only one MWD (Grw+70$^o$~8247) has a cooling
time less than 1~Gyr, and only just barely.  While the sample is
relatively small, the radius bias of the apparent-magnitude limited
samples may not be the whole story.  We acknowledge the possibility that
an unknown physical mechanism causes the fractional incidence of MWDs to
increase with decreasing $T_{\rm eff}$, luminosity and/or cooling age.
Two possible causes are discussed briefly in Valyavin \& Fabrika (1999),
who presented similar arguments about observed samples of WDs.

The total fraction of MWDs will increase more if precise
spectropolarimetric analyses capable of detecting fields down to
$\sim$10~kG are applied to samples like these.  Precise observations
need to be applied to a larger, complete sample of WDs at all
temperatures and cooling ages.  Finally, we note that the hypothesized
origin of MWDs from peculiar (Ap) stars may have to be reevaluated.
Kawka et al. (2002) explore this possibility, suggesting that normal
main sequence stars with low fields may also be required to provide 
enough progenitors.

\acknowledgments 

This work was supported in part by the NSERC Canada and by the Fund
NATEQ (Qu\'ebec). We thank Tom Marsh for the suggestion to revisit this
issue with the PG sample.  We thank the referee, John Landstreet, for 
useful comments and suggestions.

\newpage 

\begin{figure} 

\caption{The fit to the Zeeman-split, H-line spectrum of 
PG~1220+234 is shown.  The technique is applicable to field 
strengths $\la$10~MG where only linear splitting without substantial 
quadratic shifts are taking place (see BLR and references therein). 
These parameters are listed in Table~1. }

\end{figure} 

\begin{figure} 

\caption{A plot of estimated $T_{\rm eff}$ vs mass (in $M_\odot$) for the
nearby star sample discussed in section 2.3.  Filled circles are MWDs
with estimated surface field strengths $\ga 2$~MG.  Open circles are
stars not known to be strongly magnetic. Isochrones are shown for an
assumed C/O core composition, $q({\rm He})=10^{-2}$, and $q({\rm
H})=10^{-4}$ from BLR, labelled in units of $10^9$ years.  The filled
curves are for the WD cooling times alone, the dashed include the
stars' expected nuclear lifetimes.  }

\end{figure} 

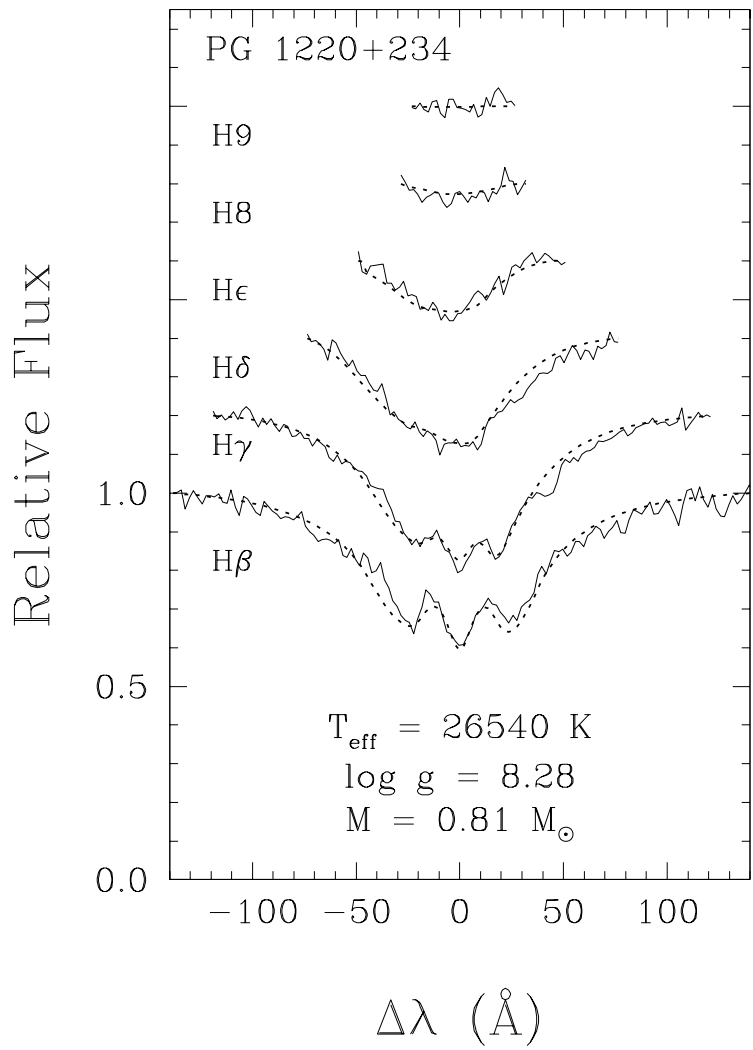
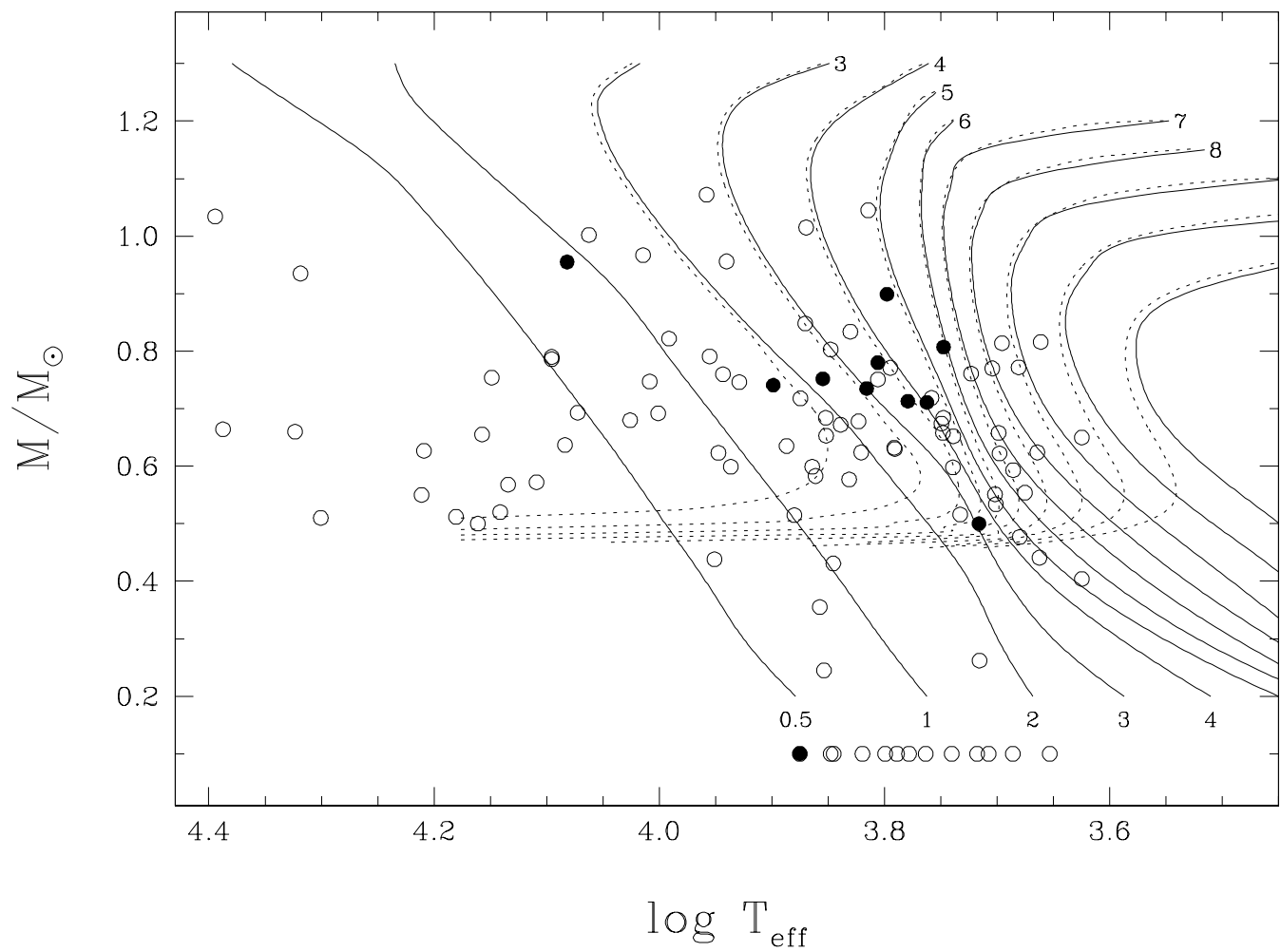
\end{document}